\documentclass[twocolumn]{aastex62}

\newcommand{\lat}{$Fermi$-LAT}

\hypersetup{linkcolor=red,citecolor=cyan,filecolor=blue,urlcolor=magenta}
\usepackage{subfigure}
\shorttitle{Tentative Fermi-LAT detection of the blazar B3 1428+422.}
\shortauthors{Liao et al.}

\begin{document}

\title{{\it Fermi}-LAT detection of a transient $\gamma$-ray source in the direction of a distant blazar B3 1428+422 at $z =4.72$}

\correspondingauthor{Neng-Hui Liao,Yi-Zhong Fan}
\email{liaonh@pmo.ac.cn, yzfan@pmo.ac.cn}

\author[0000-0001-6614-3344]{Neng-Hui Liao}
\affiliation{Key Laboratory of Dark Matter and Space Astronomy, Purple Mountain Observatory, Chinese Academy of Sciences, Nanjing 210034, China}

\author{Shang Li}
\affiliation{Key Laboratory of Dark Matter and Space Astronomy, Purple Mountain Observatory, Chinese Academy of Sciences, Nanjing 210034, China}
\affiliation{University of Chinese Academy of Sciences, Yuquan Road 19, Beijing 100049, China}


\author[0000-0002-8966-6911]{Yi-Zhong Fan}
\affiliation{Key Laboratory of Dark Matter and Space Astronomy, Purple Mountain Observatory, Chinese Academy of Sciences, Nanjing 210034, China}

\begin{abstract}
We report the detection of a transient $\gamma$-ray source in the direction of B3 1428+422 ($z=4.72$) by analyzing the 110-month \lat~{\tt Pass} 8 data. The new transient $\gamma-$ray source is far away from the Galactic plane and has a rather soft spectrum, in agreement with being a high redshift blazar.
We suggest that the newly discovered transient is the $\gamma$-ray counterpart of B3 1428+422, which could be the {\it most distant} GeV source detected so far.  The detection of a group of such distant $\gamma-$ray blazars will be helpful to reconstruct the evolution of the luminosity function and to study the extragalactic background light at such high redshifts.
\end{abstract}


\section{Introduction} \label{sec:intro}
Blazars, including flat-spectrum radio quasars (FSRQs) and BL Lacertae objects (BL Lacs), are an extreme subclass of Active Galactic Nuclei (AGNs) whose strong relativistic jets are closely aligned with our line of sight \citep{BR78,1995PASP..107..803U}. Since the jet emissions are strongly boosted due to relativistic effects,  they are the dominant population among the extragalactic $\gamma$-ray sky \citep[e.g.,][]{2015ApJS..218...23A}. At early cosmic time,  blazars serve as luminous beacons \citep[e.g.,][]{2004ApJ...610L...9R}, harboring supermassive black holes (SMBHs) heavier than one billion solar masses \citep{2010MNRAS.405..387G}. High redshift blazars not only provide crucial information of formation and growth of the first generation of SMBHs as well as their jets, but also reveal potential impacts of the jets on the evolution of AGNs along with their host galaxies \citep[e.g.,][]{2010A&ARv..18..279V,2012ARA&A..50..455F}.

Some quasars tentatively classified as blazars with redshifts of $\geq5$ have been discovered \citep{2004ApJ...610L...9R,2012MNRAS.426L..91S,2014MNRAS.440L.111G,2014ApJ...795L..29Y}. However, none of these sources have been identified as $\gamma$-ray emitters. NVSS~J151002+570243 ($z=4.3$, \citealt{2015PASA...32...10F}) is the most distant $\gamma$-ray blazar reported in the literature \citep{2017ApJ...837L...5A}. One of the major characteristics of these high redshift sources is that they have softer $\gamma$-ray spectra ($\Gamma_{\gamma} \simeq$ 3, \citealt{2017ApJ...837L...5A, 2018ApJ...853..159L}) than the nearby sources ($\Gamma_{\gamma} \simeq$ 2.5, \citealt{2015ApJ...810...14A}). Since the angular resolution of {\it Fermi}-LAT for sub-GeV photons is much worse than for GeV $\gamma$ rays\footnote{http://www.slac.stanford.edu/exp/glast/groups/canda/\\lat\_Performance.htm \label{latper}}, the detection of the high redshift soft  $\gamma$-ray sources is challenging. Interestingly, a 10-fold $\gamma$-ray flux increase and a harder-when-brighter spectral variability behavior have been detected in two $\gamma$-ray blazars beyond redshift 3 \citep{2018ApJ...853..159L}. Therefore, the very distant blazars may be relatively easier to be detected in a outburst phase.

B3 1428+422 ($z$=4.72, \citealt{1998MNRAS.294L...7H}), also known as GB 1428+4217, was identified in an optical spectroscopy search for high redshift flat-spectrum radio sources. Its high X-ray luminosity ($\rm \sim 10^{47}$ erg $\rm s^{-1}$, \citealt{1997MNRAS.291L...5F}), the hard X-ray spectrum ($\Gamma_{x} \simeq$ 1.5, \citealt{2004MNRAS.350L..67W}), the radio morphology including a compact dominant core with high brightness temperature ($T_{b} \simeq$ 5$\times 10^{11}$ K, \citealt{2010A&A...521A...6V}) and more importantly the significant radio and X-ray variability \citep{1999MNRAS.308L...6F,2006MNRAS.368..844W} strongly suggest that B3 1428+422 is a highly active high redshift blazar. Searches for its $\gamma$-ray emission with the $\sim$ 7.5 years {\it Fermi}-LAT data have been performed, but no significant signal has been identified \citep{2016ApJ...825...74P,2017ApJ...837L...5A}. In this Letter, we analyze the $\sim 9$ years {\it Fermi}-LAT data (Section \ref{sec:data}), and report a promising $\gamma$-ray counterpart of B3 1428+422 (Section \ref{sec:resul}), along with some discussions (Section \ref{sec:diss}).

\section{Data Analysis} \label{sec:data}
The first 110 months (MJD 54683$-$58032) {\tt SOURCE} type \lat~data ({\tt evclass} = 128 and {\tt evtype} = 3), in the energy range of $0.1-500$ GeV, are analyzed with the updated {\it Fermi} Science Tools package of version {\tt v11r5p3}. The entire data set is filtered with {\tt gtselect} and {\tt gtmktime} tasks, by adopting a maximum zenith angle of 90$\degr$ and ``{\tt DATA\_QUAL > 0}'' \&  ``{\tt LAT\_CONFIG==1}''. Then the {\tt unbinned} likelihood algorithm implemented in the {\tt gtlike} task is used to extract the $\gamma$-ray flux and spectrum. Since B3 1428+422 is not included in any current $\gamma$-ray catalogs, a corresponding $\gamma$-ray source located at the radio position of B3 1428+422 with a single power-law (i.e. $dN/dE \propto E^{-\Gamma}$, where $\Gamma$ is the spectral photon index) spectral template is added in the analysis model file. Meanwhile, all sources in the preliminary LAT 8-year Point Source List (FL8Y\footnote{https://fermi.gsfc.nasa.gov/ssc/data/access/lat/fl8y/}) within 15$\degr$ of the target have been taken into account. Parameters of the FL8Y sources lying within 10$\degr$ radius of interest (ROI) as well as two diffuse templates are left free, while others are fixed at FL8Y values. The test statistic (TS = 2$\rm \Delta$log$\mathcal{L}$, \citealt{1996ApJ...461..396M}) is adopted to quantify the significance of a $\gamma$-ray source, where $\mathcal{L}$ represents the likelihood function, between models with and without the source. In the extraction of $\gamma$-ray light curves, weak background sources with TS values below $10$ are removed from the model file.
If new sources emerge in the subsequently generated TS residual map with TS values higher than 25 after the likelihood analysis, they are added into the updated background model and the likelihood fitting is re-performed.

\section{Results} \label{sec:resul}
First, we perform a fit of the entire 110-month data. The TS map reveals two $\gamma$-ray sources {not included} in FL8Y, with TS values of 31 and 27, respectively. Their optimized locations are R.A.~ 215.31$\degr$ and DEC.~ 39.03 as well as R.A.~ 215.92$\degr$ and DEC.~ 40.72,  with 95\% C. L. radii of 0.15$\degr$ and 0.34$\degr$, respectively. The first source is likely associated with a radio-loud narrow-line Seyfert 1 FBQS~J142106.0+385522, as reported in \cite{2018ApJ...853L...2P}, while no promising low-energy counterpart has been found for the second source. Adopting the updated background model file including these two sources, an analysis of the entire 110-month data are carried out again. There is no evidence for a bright $\gamma$-ray source in the direction of B3 1428+422. The TS value of the potential weak $\gamma$-ray source ($\Gamma$ fixed to 3.0) is rather small (TS$=3$), in agreement with the result of a previous study \citep{2016ApJ...825...74P}. This result holds when the photon index is fixed as 2.6 and 2.8 instead, according to the photon indexes of high redshift blazars detected by {\it Fermi}-LAT \citep{2017ApJ...837L...5A}.

\begin{figure}[ht!]
\centering
\subfigure[]{%
  \includegraphics[width=0.45\textwidth]{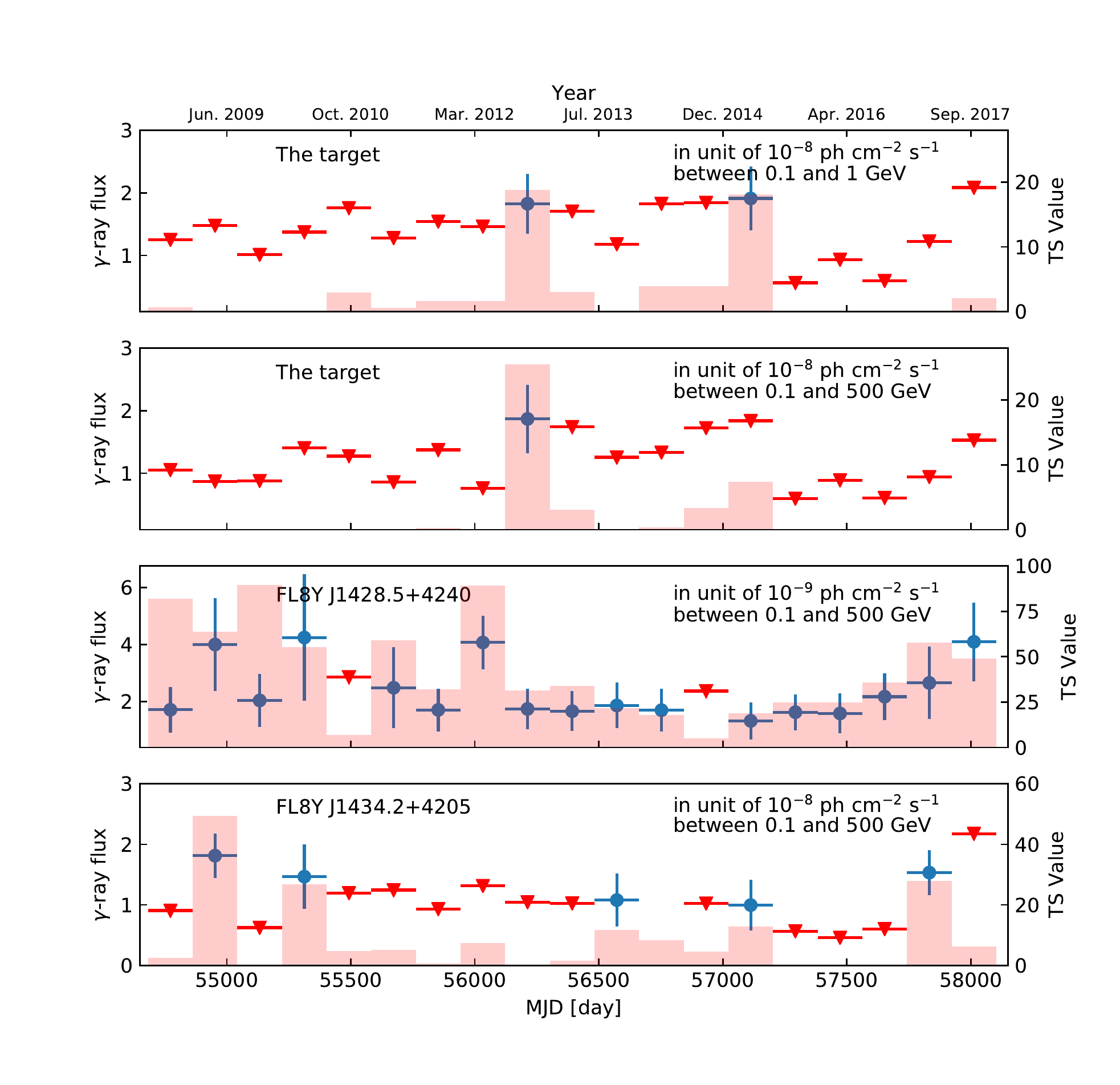}
  \label{fig:hylc}}
\subfigure[]{%
  \includegraphics[width=0.45\textwidth]{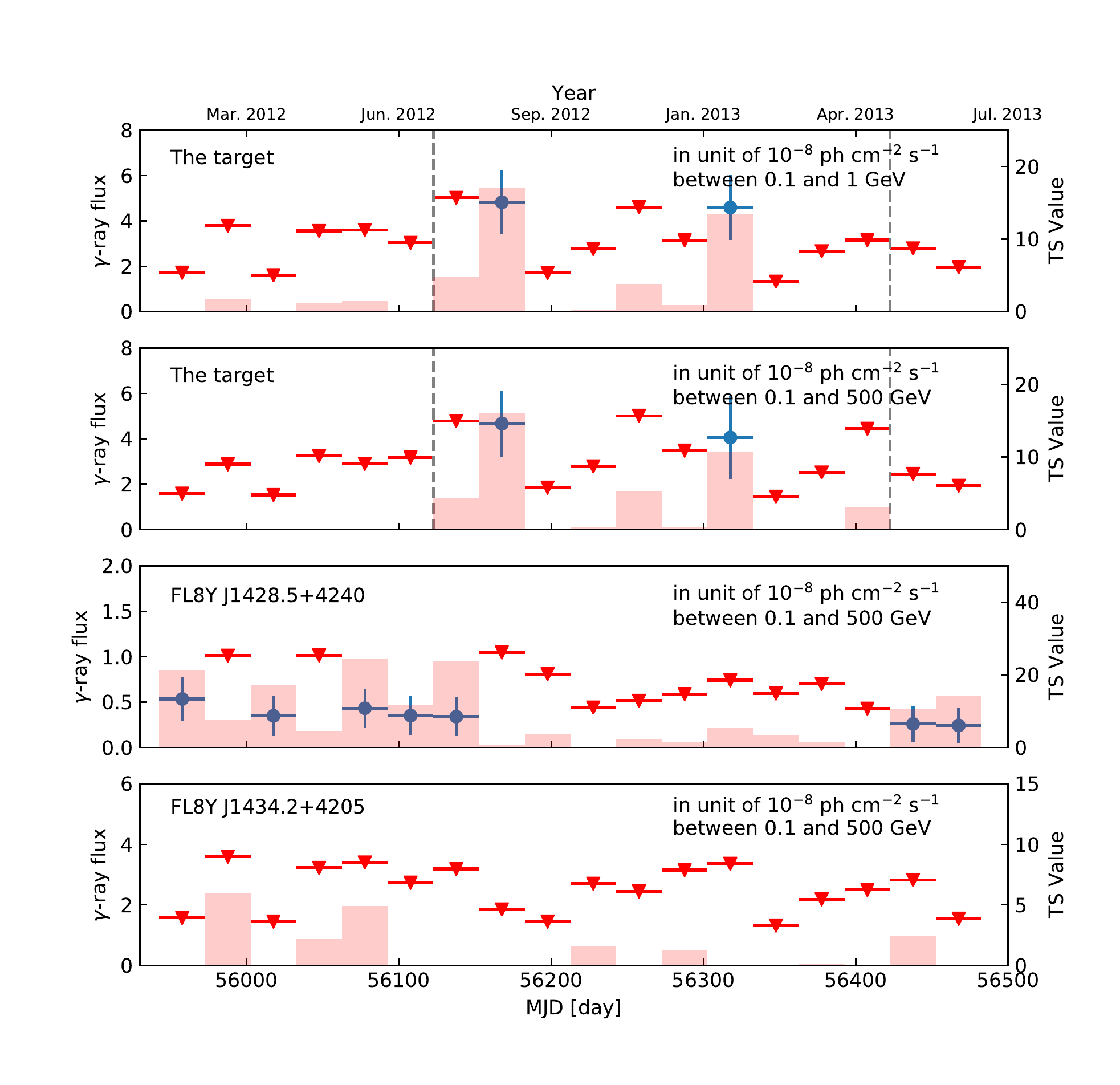}
  \label{fig:mlc}}
\caption{Half-year bin (a) and monthly (b) $\gamma$-ray light curves of the target as well as its neighbor FL8Y~J1428.5+4240 and FL8Y~J1434.2+4205. Blue points represent the $\gamma$-ray fluxes, while the red triangles are upper limits. Red bars are the corresponding TS values. The grey dashed vertical lines mark the flaring epoch of the transient source.
\label{fig:f1}}
\end{figure}

\begin{figure}[ht!]
\centering
\subfigure[Pre-flare epoch]{%
  \includegraphics[width=0.3\textwidth]{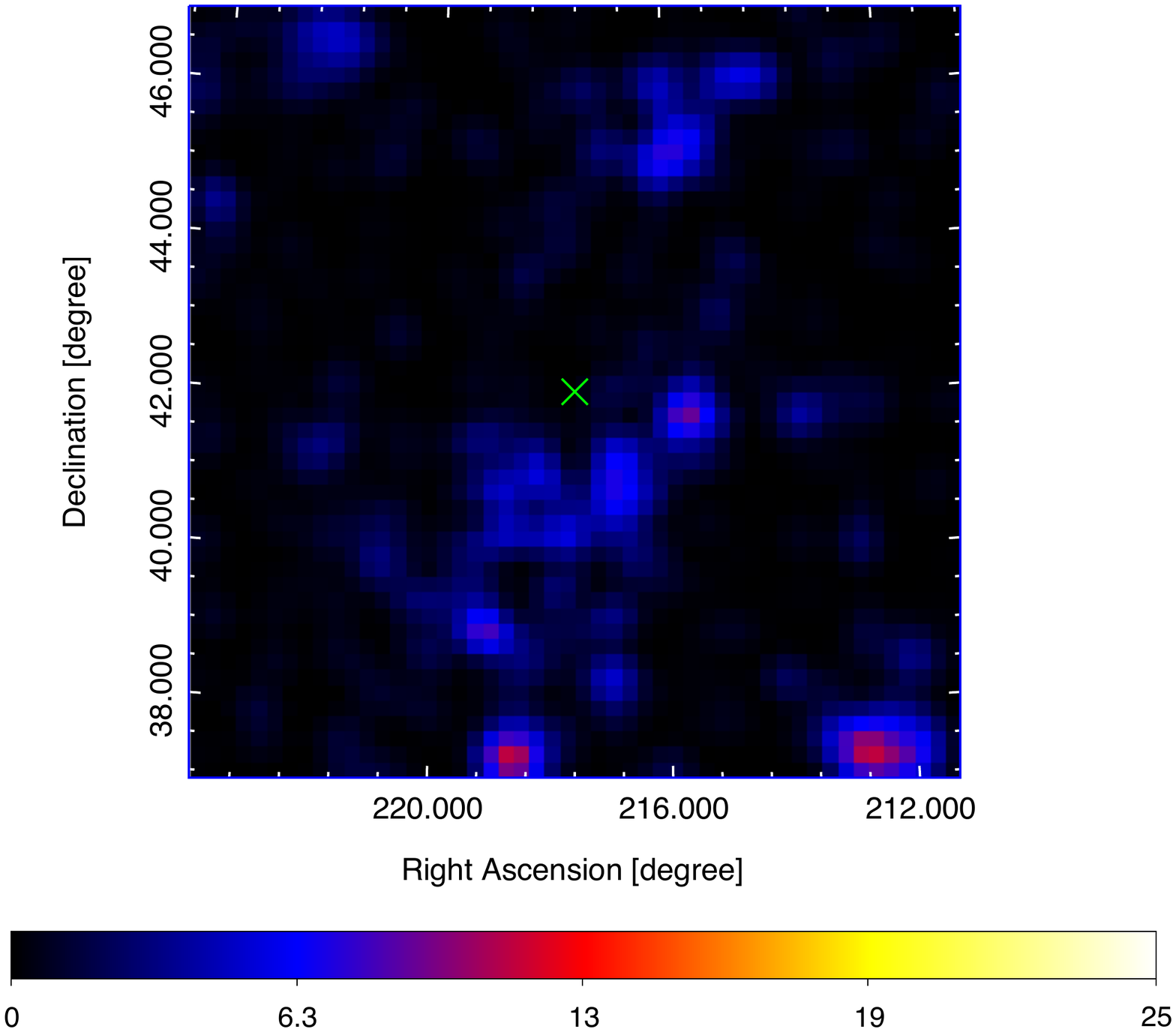}
  \label{fig:subfigure1}}
\quad
\subfigure[Flare-epoch]{%
  \includegraphics[width=0.30\textwidth]{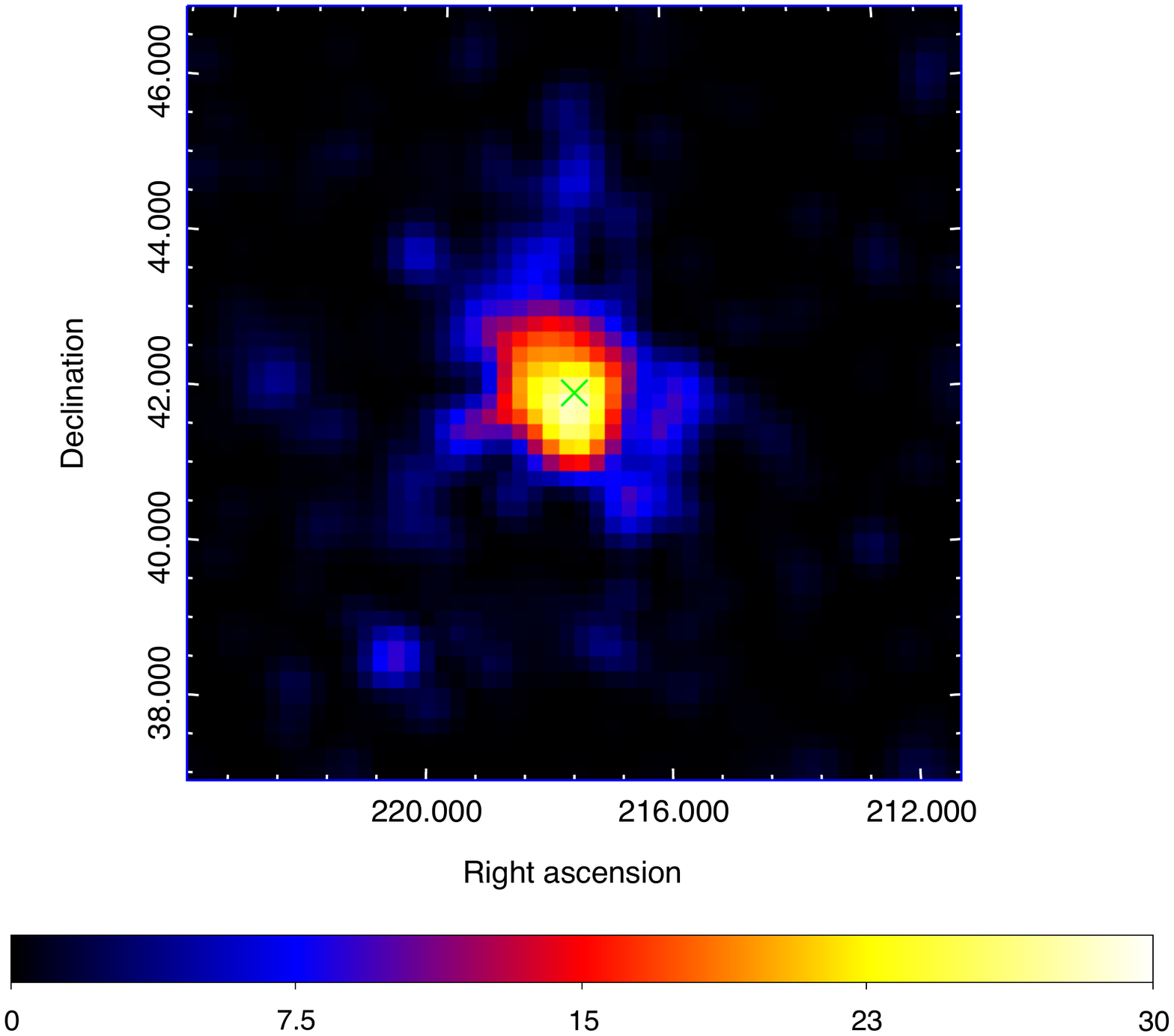}
  \label{fig:subfigure2}}
\quad
\subfigure[Post-flare epoch]{%
  \includegraphics[width=0.3\textwidth]{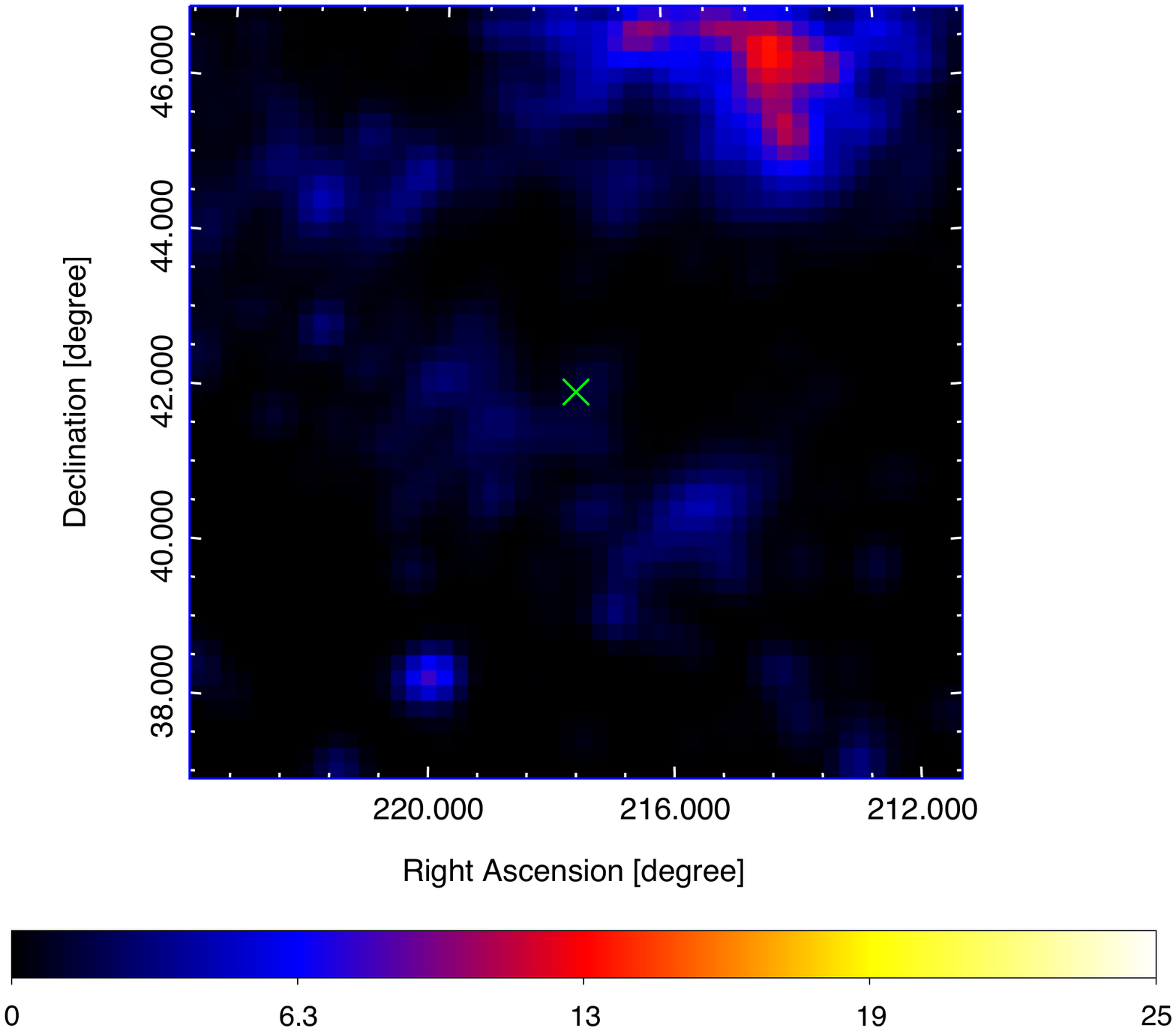}
  \label{fig:subfigure3}}

\subfigure[Zoomed-in view of the flare-epoch]{%
  \includegraphics[width=0.4\textwidth]{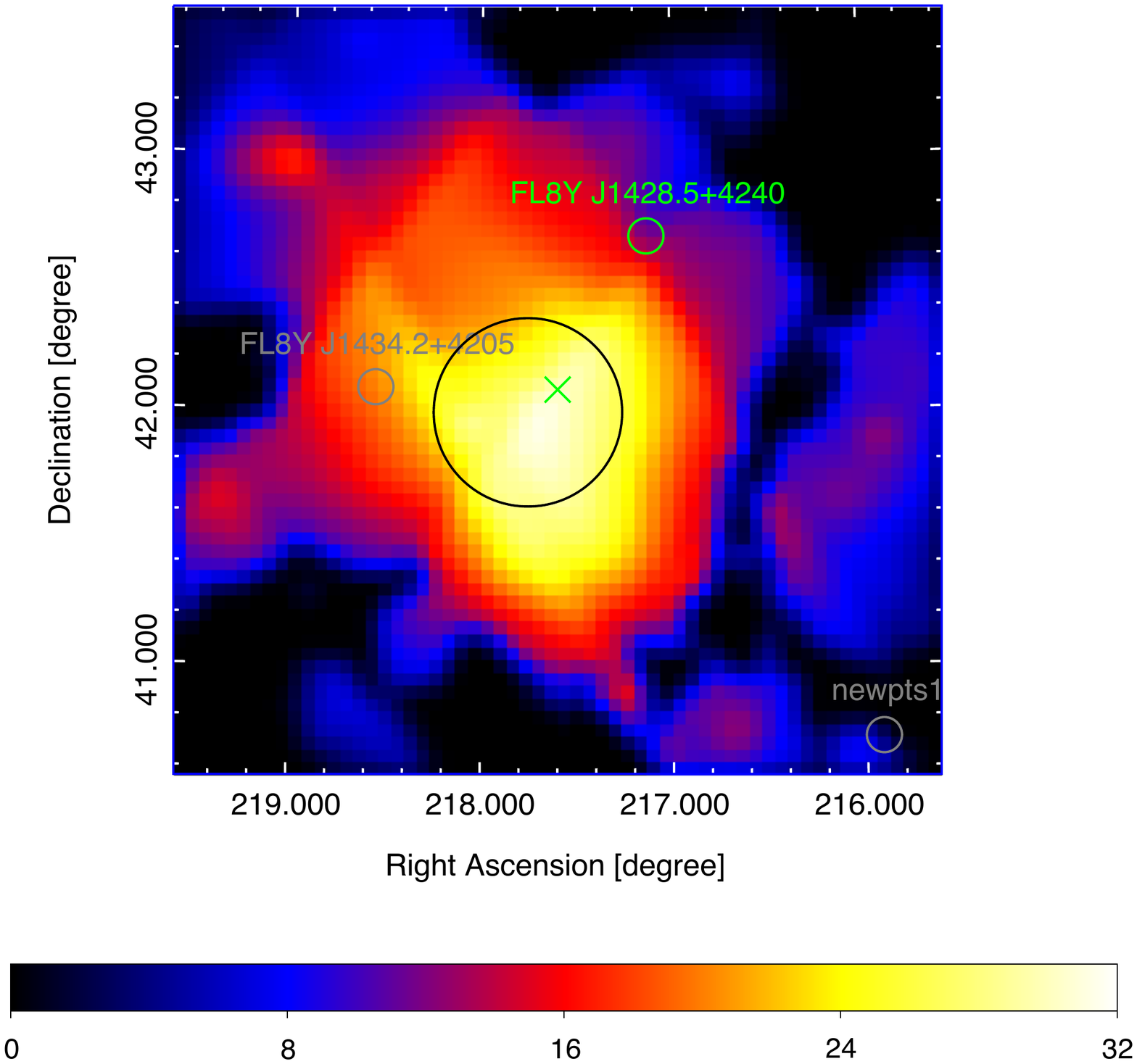}
  \label{fig:subfigure4}}
\quad
\caption{The TS maps for different epochs. The upper three panels represent 10$\degr \times$10$\degr$ TS maps with 0.2$\degr$ per pixel centered at B3 1428+422 between 0.1 and 500~GeV for the pre-flare epoch (a), the flare epoch (b) and post-flare epoch (c), respectively. The panel (d) is a zoomed-in 3$\degr \times$3$\degr$ view of the panel (b) with 0.05$\degr$ per pixel centered at its optimized $\gamma$-ray location. The green X-shaped symbol represents the radio position of B3 1428+422. The black circle in panel (d) is the 95\% C. L. error radii of the locations of the $\gamma$-ray source. Locations of the nearby background sources considered in the analysis of entire time range of LAT data are also marked. In fact, only FL8Y J1428.5+4240 (in color green) remains in the model file, whereas neighbors (in color grey) have been removed due to their low TS values.}
\label{fig:f2}
\end{figure}

We then extract a half-year bin $\gamma$-ray light curve. When one source is not significantly detected by \lat~(i.e. TS $<$ 10), a 95\% confidential level (C. L.) upper limit is obtained by {\tt pyLikelihood UpperLimits} tool. Though the TS values in most time bins are rather small ($<$ 4), the 9th bin (MJD 56123$-$56303, i.e. from July 2012 15th to Jan. 2013 11th) is distinguished by a high TS value of 26, which reveals the emergence of a new $\gamma$-ray transient source (see Figure \ref{fig:hylc}). Due to the limited spatial resolution of \lat, such a rise of TS value could be caused by a flaring bright neighbor \citep[e.g.,][]{2018ApJ...853..159L}. There are two known $\gamma$-ray sources nearby, FL8Y J1428.5+4240 (0.9$\degr$ away) and FL8Y J1434.2+4205 (0.6$\degr$ away), listed in FL8Y and associated with H~1426+428 and B3~1432+422, respectively. Their temporal behaviors are also examined. As shown in Figure \ref{fig:hylc}, the appearance of the new $\gamma$-ray source does not coincide with any flaring event of its neighbors. Further 1-month time bin light curve is extracted to identify the exact flaring epoch, see Figure \ref{fig:mlc}. A period of 10 months is marked, in the time range between MJD 56123 and MJD 56423 (i.e. from July 2012 15th to May 2013 11th) Analyses for the pre-flare, flare and post-flare epochs are performed. No significant signals are found in the pre-flare and post-flare phases, see Figure \ref{fig:subfigure1} and \ref{fig:subfigure3}. However, a strong $\gamma$-ray signal indeed appears at the direction of B3 1428+422, see the corresponding TS maps in Figure \ref{fig:subfigure2}, confirming the result of the half-year bin $\gamma$-ray light curve.

A localization analysis of the new $\gamma$-ray source during the 10-month period provides the coordinates of R.A. 217.75$\degr$ and DEC. 41.99$\degr$, with a 95\% C. L. error radius of 0.33$\degr$. The angular separation between the $\gamma$-ray position and the radio position of B3 1428+422 is 0.14$\degr$, see Figure \ref{fig:subfigure4}. We have also looked for other potential counterparts, especially blazar candidates included in the BZCAT list \citep{2009A&A...495..691M} and high frequency radio surveys \citep[e.g.,][]{2003MNRAS.341....1M,2007ApJS..171...61H,2008ApJS..175...97H}. No other sources in these catalogs are found to be within the 95\% C.L. $\gamma$-ray error radius. We use the Bayesian association method as well as the corresponding prior probability value for CRATES catalog (0.33, \citealt{2010ApJS..188..405A}) to calculate the association probability. Our result is 0.81, above the threshold of $0.8$, suggesting a likely association. Adopting the updated $\gamma$-ray position, a single power-law function provides an acceptable description of the $\gamma$-ray spectrum of the source,
\begin{equation}
 \frac{dN}{dE}=(2.16\pm0.44)\times10^{-13}(\frac{E}{\rm 263~MeV})^{-(2.95\pm0.24)},
\end{equation}
and the photon flux is $\rm (1.92\pm0.45) \times 10^{-8}$ ph $\rm cm^{-2}$ $\rm s^{-1}$. If the $\gamma$-ray source is indeed associated with B3 1428+422, the apparent isotropic $\gamma$-ray luminosity in the flare phase should be $(8\pm2) \times 10^{48}$ erg $\rm s^{-1}$ (here we take a $\Lambda$CDM cosmology with $H_{0}=67~{\rm km~ s^{-1}~Mpc^{-1}}$, $\Omega_{\rm m}=0.32$, and $\Omega_{\Lambda}=0.68$; \citealt{2014A&A...571A..16P}). The corresponding TS value is 32 ($\simeq 4.8 \sigma$). Since we have 11 trials and hence a global significance after correction of a trial factor can be estimated as
\begin{equation}
CDF(\chi^{2}_{dof=4};TS_{max})^{11} = CDF(\chi^{2}_{dof=1};\sigma^{2}),
\end{equation}
which is still $> 4\sigma$. By comparison with the 110~months averaged flux status, a 95\% C. L. upper limit of $5 \times 10^{-9}$ ph $\rm cm^{-2}$ $\rm s^{-1}$, there is a significant $\gamma$-ray flux increase, though no explicit variability amplitude can be obtained due to the limited statistic.

\begin{figure}[ht!]
\plotone{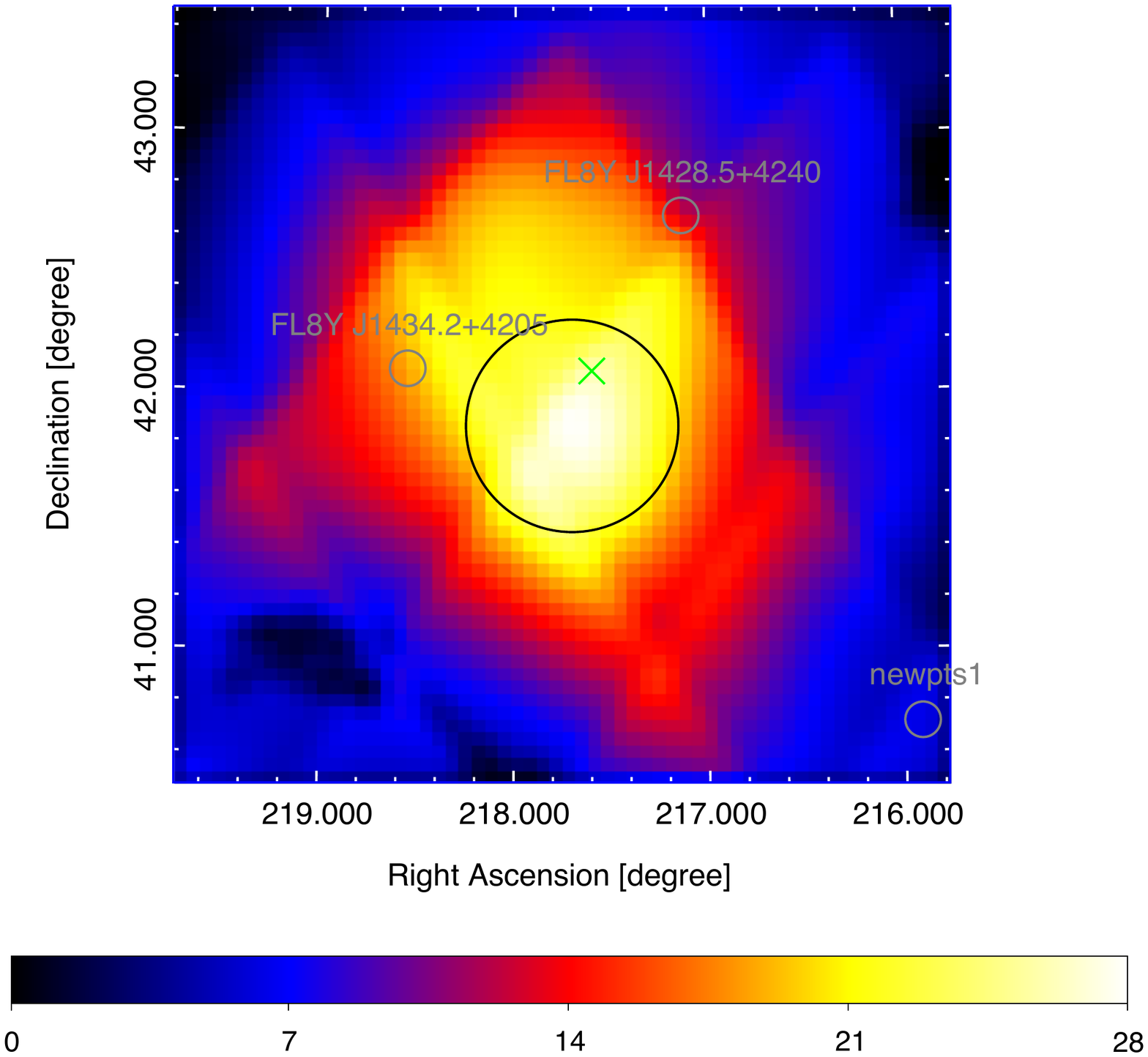}
\caption{The TS map as same as \ref{fig:subfigure4} but in the 0.1-1 GeV energy range. Note that all background sources nearby (in color grey) now have been removed  due to their low TS values\label{fig:f3}}
\end{figure}

\begin{figure}[ht!]
\plotone{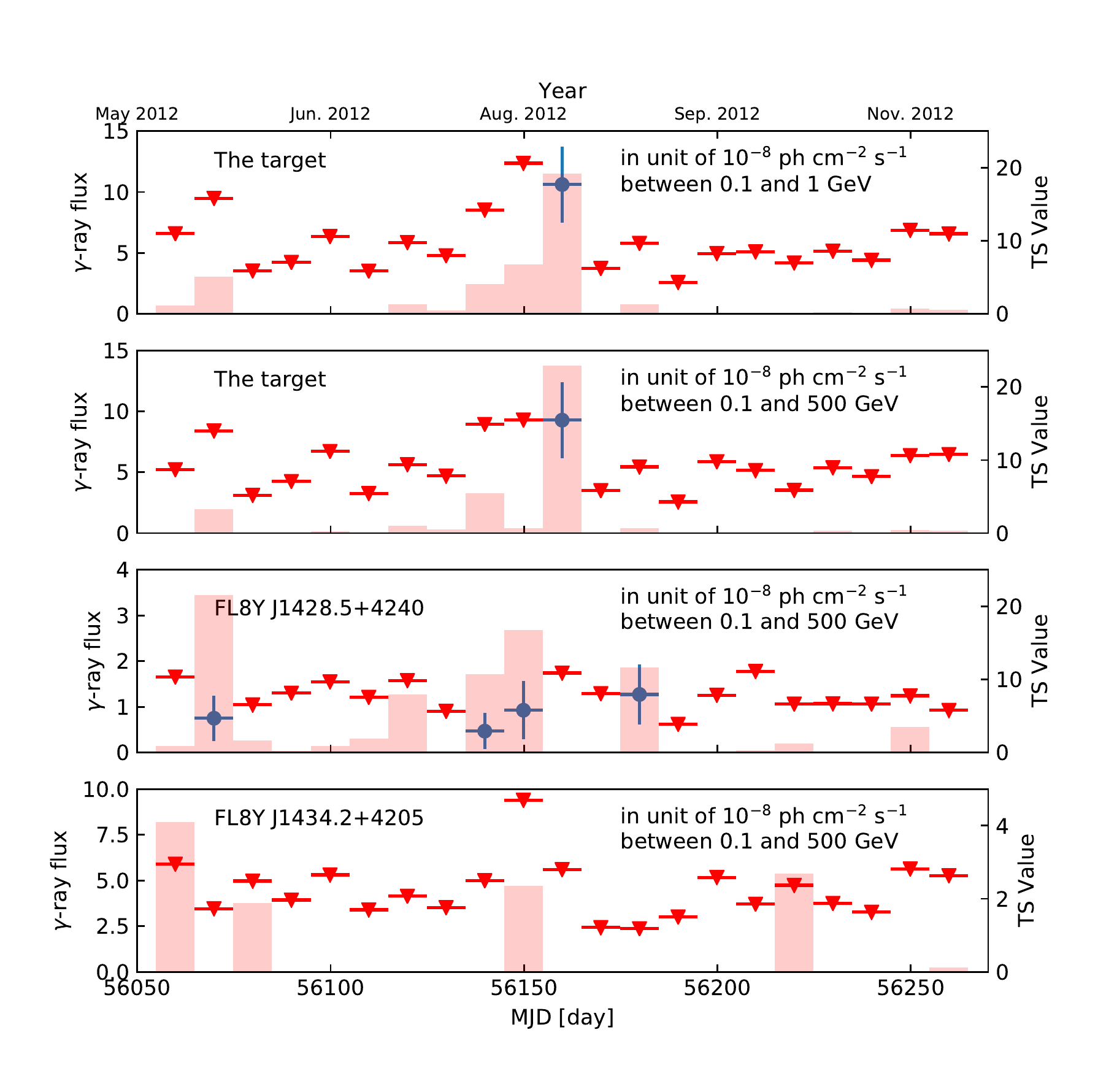}
\caption{The 10-day time bin $\gamma$-ray light curves of the target as well as its neighbor FL8Y~J1428.5+4240 and FL8Y~J1434.2+4205. Blue points represent the $\gamma$-ray fluxes, while the red triangles are upper limits. Red bars are the corresponding TS values.
\label{fig:f4}}
\end{figure}

Unlike FL8Y J1434.2+4205, the other nearby neighbor FL8Y J1428.5+4240 is significant (TS$=42$) in the 10-month flaring epoch. Several tests are performed to evaluate its influence on our detection. No significant differences in the results obtained by using the model with one source or both sources are found. Moreover, since these two sources exhibit rather different spectral behaviors ($\Gamma_{neighbor}\simeq1.5$ while $\Gamma_{target}\simeq3.0$), the {\it Fermi}-LAT data below and above 1~GeV are analyzed separately. In the former case, the target is significant against the background (TS = 28) while the neighbor turns to be undetectable (TS $<$ 4), as shown in the TS map (see Figure \ref{fig:f3}). The photon flux between 0.1 and 1~GeV is $\rm (1.78\pm0.50) \times 10^{-8}$ ph $\rm cm^{-2}$ $\rm s^{-1}$ and the corresponding apparent luminosity is $(7\pm2) \times 10^{48}$ erg $\rm s^{-1}$ assuming a redshift of 4.72. The optimized location is R.A. 217.70$\degr$ and DEC. 41.87$\degr$, with the 95\% C.L. $\gamma$-ray error radius of 0.41$\degr$. Since the angular separation between the radio position of B3 1428+422 and the $\gamma$-ray location is 0.22$\degr$, the spatial association is confirmed. On the other hand, TS of the target becomes rather low (TS $\simeq4$) in case of analysis in the 1 $-$ 500 GeV energy range, which is in agreement with the rather soft spectrum of the target. In the 14th time bin of the half-year light curve the TS value of the target increased from 8 to 18 using the entire energy range or only the 0.1-1 GeV one, suggesting a mild activity in that period. Meanwhile, further 10-day light curves are extracted, see Figure \ref{fig:f4}. In one time bin (MJD 56155 $-$ 56165, i.e. from Aug. 2012 15th to Aug. 2012 25th), the fit results in a TS = 23 (between 0.1 and 500~GeV) or TS=18 (between 0.1 and 1~GeV), whereas the TS of the neighbor is lower than 1. Therefore, the new $\gamma$-ray source is robust rather than artificial caused by the neighbor FL8Y J1428.5+4240. In addition, localization analysis of this bin for the target gives an optimized location of R.A. 217.57$\degr$ and DEC. 41.66$\degr$, with 95\% C.L. $\gamma$-ray error radius of 0.47$\degr$, suggesting B3 1428+422 is still within the error radius.

In consideration of the soft spectrum of the new $\gamma$-ray source, it should be still detectable with LAT data with a lower energy threshold (i. e. $>$ 60~MeV instead of $>$ 100~MeV). Following the data analysis procedure adopted in \cite{2017ApJ...837L...5A}, which is a summed likelihood analysis performed by the {\tt Fermipy} software \citep{2017arXiv170709551W} based on 4 different PSF event types with 15$\degr$ ROI, customized zenith cut for each dataset as well as handled energy dispersion, we indeed find a significant $\gamma$-ray source (TS = 35) there for the 10-month flaring epoch. Adopting a single power-law as spectral template, the derived spectral index (2.95$\pm0.19$) is identical with what we find from the $>$ 100~MeV data. The corresponding photon flux between 60~MeV and 500~GeV is $\rm (4.5\pm1.0) \times 10^{-8}$ ph $\rm cm^{-2}$ $\rm s^{-1}$. A localization analysis provides an optimized location of R.A. 217.53$\degr$ and DEC. 41.66$\degr$ with a $95\%$ C.L. $\gamma$-ray error radius of 0.57$\degr$, from which B3 1428+422 remains to be within the error radius.

\section{Summary and Discussion} \label{sec:diss}
Highly variable $\gamma$-ray emissions from high redshift blazars (i.e. $z >$ 2) have been detected by \lat \citep{2013A&A...556A..71A,2014MNRAS.444.3040O,2015ApJ...799..143A,2016ApJ...825...74P,2016MNRAS.455.1881D,2018ApJ...853..159L}. It is reasonable to record such violent behaviors because significantly beamed bright sources are preferred to be detected there due to the Malmquist bias. The typical peak $\gamma$-ray luminosity of these high redshift sources is $\sim 10^{50}$ erg $\rm s^{-1}$ \citep[e.g.,][]{2018ApJ...853..159L}. PKS 1830$-$211 ($z$ = 2.5) is the brightest high redshift blazar detected so far, which has a daily peak flux of 3$\times 10^{50}$ erg $\rm s^{-1}$ \citep{2015ApJ...799..143A}.  By comparison, the 10-day peaking luminosity of our target is $\sim 3 \times 10^{49}$ erg $\rm s^{-1}$ assuming a  redshift of 4.72. Together with its soft $\gamma$-ray spectrum, the robustness of the $\gamma$-ray signal and the spatial association between the $\gamma$-ray source and B3 1428+422, we suggest that the new transient GeV source may be the $\gamma$-ray counterpart of B3 1428+422.

High redshift $\gamma$-ray sources, including blazars and $\gamma$-ray bursts (GRBs), are valuable targets because of the imprints of extragalactic background light (EBL) in their $\gamma$-ray spectra. So far the most distant GRB detected by \lat~is GRB 080916C at a photometric redshift of 4.35 \citep{2009A&A...498...89G,2009Sci...323.1688A}. Therefore B3 1428+422 could be the farthest high energy $\gamma$-ray source detected so far. Horizon $\gamma$-ray photons (i.e. suffered significant EBL attenuation, $\tau_{\gamma\gamma}$ = 1) have been detected in several blazars \citep[e.g.,][]{2013ApJ...777L..18T}. Though energy of the most energetic $\gamma$-ray photons from our transient is about 2~GeV which could not challenge the current EBL models ($\rm E_{horizon}\sim30~GeV$, \citealt{2010ApJ...712..238F}), the presence of GeV sources at $z>4.5$ is indeed encouraging for such a purpose. Moreover, since there are no known $\gamma$-ray BL Lacs beyond redshift 3 right now \citep{2015RAA....15..313L}, detections of high redshift $\gamma$-ray FSRQs is also crucial to determine the high redshift end of $\gamma$-ray luminosity function (GLF) of blazars. In fact, PKS 0537$-$286 ($z$ = 3.1) remains to be the most distant $\gamma$-ray source among the sample used to generate the current blazar GLFs \citep[e.g.,][]{2012ApJ...751..108A,2013MNRAS.431..997Z}. Therefore, the recently detected new high redshift $\gamma$-ray FSRQs, especially the five new sources with redshifts between 3.4 and 4.3 \citep{2017ApJ...837L...5A} as well as B3 1428+422, should be embraced to update the blazar GLF (Liao \& Zeng in prep.).

Compared with other high redshift $\gamma$-ray blazars \citep{2017ApJ...837L...5A}, the error radius of the $\gamma$-ray source tentatively
associated with B3 1428+422 is relatively large due to its faintness and more importantly the rather soft spectrum, especially for the analyses of the 10-day time bin and the 10-month period with $>$ 60 MeV LAT data, making the association between the $\gamma$-ray source and the low-energy counterpart more difficult. Nevertheless, it is still helpful to examine whether there are only other suitable low-energy counterparts in the ``enlarged" $\gamma$-ray error radii. Except B3 1428+422, we found no suitable blazar candidates in the radio catalogs cited in the previous Section. The same holds if we choose the blazar candidates from the WISE blazar-like radio-loud sources \citep{2014ApJS..215...14D} instead. We also notice that B3 1428+422 is the only flat-spectrum radio source in the region of the interest. However, for some of the radio sources included in the LAT error circle, only NVSS images are available and it is not possible to determine a spectrum and thus if these are flat spectrum radio sources or not.


Multiwavelength campaigns, including $\gamma$-ray observation as well as complementary observations from radio to X rays, become a routine approach to probe the physical processes of AGN jets, and simultaneous $\gamma$-ray and optical flares have been frequently detected for FSRQs \citep[e.g.,][]{2010Natur.463..919A}. The simultaneous detection of these flares would provide a decisive proof of the association between the $\gamma$-ray source and its optical counterpart \citep[e.g.,][]{2016ApJS..226...17L}. Such an important proof is however still lacking for all known $\gamma$-ray FSRQs beyond redshift 3. Due to their faint optical emission (typically $\rm R_{mag} \ga$ 20), many optical transient surveys are not deep enough \citep[e.g.,][]{2009ApJ...696..870D}. Nevertheless, strong optical flares are found from one of those sources, NVSS J163547+362930 ($z$ = 3.6), based on archival Palomar Transient Factory (PTF) data \citep{2018ApJ...853..159L}. Therefore, we examined the PTF data\footnote{http://irsa.ipac.caltech.edu/applications/ptf/} of B3 1428+422 ($\rm r_{SDSS,mag} \simeq $ 21, \citealt{2014A&A...563A..54P}). No PTF observations are available in August 2012 when the intense $\gamma$-ray flare appeared and no significant optical flare of B3 1428+422 can be identified at other times. A confirmed association with the lower-energy counterpart will be achieved with future simultaneous multi-frequency observations in the period of the high $\gamma$-ray activity of the new transient. With upcoming wide-deep-fast sky survey facilities, such as the Large Synoptic Survey Telescope \citep{2008arXiv0805.2366I} as well as other future observational facilities in time domain (e.g. the Wide-Field InfraRed Survey Telescope, \citealt{2012arXiv1208.4012G}; the Einstein Probe, \citealt{2015arXiv150607735Y}), a comprehensive broadband dynamic view of high redshift $\gamma$-ray sources will be achieved.

\acknowledgments
We appreciate the instructive suggestions from the anonymous referee that led to a substantial improvement of this work. Vaidehi Paliya is appreciated for providing procedures of analysis of $>$ 60~MeV LAT data in their study. This research has made use of data obtained from the High Energy Astrophysics Science Archive Research Center (HEASARC), provided by $\rm NASA^{\prime}$s Goddard Space Flight Center. This research has also made use of the NASA/IPAC Extragalactic Database and the NASA/IPAC Infrared Science Archive which are operated by the Jet Propulsion Laboratory, California Institute of Technology, under contract with the National Aeronautics and Space Administration. This research makes use of the SIMBAD database, operated at CDS, Strasbourg, France.

This work was supported in part by the National Basic Research Program of China (No. 2013CB837000), NSFC under grants 11525313 (i.e., Funds for Distinguished Young Scholars) and 11703093.

\vspace{5mm}
\facilities{{\it Fermi} (LAT)}

\clearpage

\begin{thebibliography}{}
\bibitem[Abdo et al.(2009)]{2009Sci...323.1688A} Abdo, A.~A., Ackermann, M., Arimoto, M., et al.\ 2009, Science, 323, 1688
\bibitem[Abdo et al.(2010a)]{2010ApJ...723.1082A} Abdo, A.~A., Ackermann, M., Ajello, M., et al.\ 2010a, \apj, 723, 1082
\bibitem[Abdo et al.(2010b)]{2010Natur.463..919A} Abdo, A.~A., Ackermann, M., Ajello, M., et al.\ 2010b, \nat, 463, 919 
\bibitem[Abdo et al.(2010)]{2010ApJS..188..405A} Abdo, A.~A., Ackermann, M., Ajello, M., et al.\ 2010c, \apjs, 188, 405
\bibitem[Abdo et al.(2015)]{2015ApJ...799..143A} Abdo, A.~A., Ackermann, M., Ajello, M., et al.\ 2015, \apj, 799, 143
\bibitem[Acero et al.(2015)]{2015ApJS..218...23A} Acero, F., Ackermann, M., Ajello, M., et al.\ 2015, \apjs, 218, 23
\bibitem[Ackermann et al.(2015)]{2015ApJ...810...14A} Ackermann, M., Ajello, M., Atwood, W.~B., et al.\ 2015, \apj, 810, 14
\bibitem[Ackermann et al.(2017)]{2017ApJ...837L...5A} Ackermann, M., Ajello, M., Baldini, L., et al.\ 2017, \apjl, 837, L5
\bibitem[Ajello et al.(2012)]{2012ApJ...751..108A} Ajello, M., Shaw, M.~S., Romani, R.~W., et al.\ 2012, \apj, 751, 108
\bibitem[Akyuz et al.(2013)]{2013A&A...556A..71A} Akyuz, A., Thompson, D.~J., Donato, D., et al.\ 2013, \aap, 556, A71
\bibitem[D'Abrusco et al.(2014)]{2014ApJS..215...14D} D'Abrusco, R., Massaro, F., Paggi, A., et al.\ 2014, \apjs, 215, 14
\bibitem[D'Ammando \& Orienti(2016)]{2016MNRAS.455.1881D} D'Ammando, F., \& Orienti, M.\ 2016, \mnras, 455, 1881
\bibitem[Drake et al.(2009)]{2009ApJ...696..870D} Drake, A.~J., Djorgovski, S.~G., Mahabal, A., et al.\ 2009, \apj, 696, 870
\bibitem[Orienti et al.(2014)]{2014MNRAS.444.3040O} Orienti, M., D'Ammando, F., Giroletti, M., et al.\ 2014, \mnras, 444, 3040
\bibitem[Blandford \& Rees (1978)]{BR78}Blandford, R. D., \& Rees, M. J. 1978, in Pittsburgh Conference on BL Lac Objects, ed. A. M. Wolfe (Pittsburgh, PA: Univ. Pittsburgh Press), 328
\bibitem[Fabian et al.(1997)]{1997MNRAS.291L...5F} Fabian, A.~C., Brandt, W.~N., McMahon, R.~G., \& Hook, I.~M.\ 1997, \mnras, 291, L5
\bibitem[Fabian et al.(1999)]{1999MNRAS.308L...6F} Fabian, A.~C., Celotti, A., Pooley, G., et al.\ 1999, \mnras, 308, L6
\bibitem[Fabian(2012)]{2012ARA&A..50..455F} Fabian, A.~C.\ 2012, \araa, 50, 455
\bibitem[Finke et al.(2010)]{2010ApJ...712..238F} Finke, J.~D., Razzaque, S., \& Dermer, C.~D.\ 2010, \apj, 712, 238
\bibitem[Flesch(2015)]{2015PASA...32...10F} Flesch, E.~W.\ 2015, PASA, 32, e010
\bibitem[Ghisellini et al.(2010)]{2010MNRAS.405..387G} Ghisellini, G., Della Ceca, R., Volonteri, M., et al.\ 2010, \mnras, 405, 387
\bibitem[Ghisellini et al.(2014)]{2014MNRAS.440L.111G} Ghisellini, G., Sbarrato, T., Tagliaferri, G., et al.\ 2014, \mnras, 440, L111
\bibitem[Green et al.(2012)]{2012arXiv1208.4012G} Green, J., Schechter, P., Baltay, C., et al.\ 2012, arXiv:1208.4012
\bibitem[Greiner et al.(2009)]{2009A&A...498...89G} Greiner, J., Clemens, C., Kr{\"u}hler, T., et al.\ 2009, \aap, 498, 89
\bibitem[Healey et al.(2007)]{2007ApJS..171...61H} Healey, S.~E., Romani, R.~W., Taylor, G.~B., et al.\ 2007, \apjs, 171, 61
\bibitem[Healey et al.(2008)]{2008ApJS..175...97H} Healey, S.~E., Romani, R.~W., Cotter, G., et al.\ 2008, \apjs, 175, 97
\bibitem[Hook \& McMahon(1998)]{1998MNRAS.294L...7H} Hook, I.~M., \& McMahon, R.~G.\ 1998, \mnras, 294, L7
\bibitem[Ivezic et al.(2008)]{2008arXiv0805.2366I} Ivezic, Z., Tyson, J.~A., Abel, B., et al.\ 2008, arXiv:0805.2366
\bibitem[Li et al.(2018)]{2018ApJ...853..159L} Li, S., Xia, Z.-Q., Liang, Y.-F., Liao, N.-H., \& Fan, Y.-Z.\ 2018, \apj, 853, 159
\bibitem[Liao et al.(2015)]{2015RAA....15..313L} Liao, N.-H., Bai, J.-M., Wang, J.-G., et al.\ 2015, Research in Astronomy and Astrophysics, 15, 313
\bibitem[Liao et al.(2016)]{2016ApJS..226...17L} Liao, N.-H., Xin, Y.-L., Fan, X.-L., et al.\ 2016, \apjs, 226, 17
\bibitem[Massaro et al.(2009)]{2009A&A...495..691M} Massaro, E., Giommi, P., Leto, C., et al.\ 2009, \aap, 495, 691
\bibitem[Mattox et al.(1996)]{1996ApJ...461..396M} Mattox, J.~R., Bertsch, D.~L., Chiang, J., et al.\ 1996, \apj, 461, 396
\bibitem[Myers et al.(2003)]{2003MNRAS.341....1M} Myers, S.~T., Jackson, N.~J., Browne, I.~W.~A., et al.\ 2003, \mnras, 341, 1
\bibitem[Paliya et al.(2016)]{2016ApJ...825...74P} Paliya, V.~S., Parker, M.~L., Fabian, A.~C., \& Stalin, C.~S.\ 2016, \apj, 825, 74
\bibitem[Paliya et al.(2018)]{2018ApJ...853L...2P} Paliya, V.~S., Ajello, M., Rakshit, S., et al.\ 2018, \apjl, 853, L2
\bibitem[P{\^a}ris et al.(2014)]{2014A&A...563A..54P} P{\^a}ris, I., Petitjean, P., Aubourg, {\'E}., et al.\ 2014, \aap, 563, A54
\bibitem[Planck Collaboration et al.(2014)]{2014A&A...571A..16P} Planck Collaboration, Ade, P.~A.~R., Aghanim, N., et al.\ 2014, \aap, 571, A16
\bibitem[Romani et al.(2004)]{2004ApJ...610L...9R} Romani, R.~W., Sowards-Emmerd, D., Greenhill, L., \& Michelson, P.\ 2004, \apjl, 610, L9
\bibitem[Sbarrato et al.(2012)]{2012MNRAS.426L..91S} Sbarrato, T., Ghisellini, G., Nardini, M., et al.\ 2012, \mnras, 426, L91
\bibitem[Tanaka et al.(2013)]{2013ApJ...777L..18T} Tanaka, Y.~T., Cheung, C.~C., Inoue, Y., et al.\ 2013, \apjl, 777, L18
\bibitem[Urry \& Padovani(1995)]{1995PASP..107..803U} Urry, C.~M., \& Padovani, P.\ 1995, \pasp, 107, 803
\bibitem[Veres et al.(2010)]{2010A&A...521A...6V} Veres, P., Frey, S., Paragi, Z., \& Gurvits, L.~I.\ 2010, \aap, 521, A6
\bibitem[Volonteri(2010)]{2010A&ARv..18..279V} Volonteri, M.\ 2010, \aapr, 18, 279
\bibitem[Wood et al.(2017)]{2017arXiv170709551W} Wood, M., Caputo, R., Charles, E., et al.\ 2017, arXiv:1707.09551
\bibitem[Worsley et al.(2004)]{2004MNRAS.350L..67W} Worsley, M.~A., Fabian, A.~C., Celotti, A., \& Iwasawa, K.\ 2004, \mnras, 350, L67
\bibitem[Worsley et al.(2006)]{2006MNRAS.368..844W} Worsley, M.~A., Fabian, A.~C., Pooley, G.~G., \& Chandler, C.~J.\ 2006, \mnras, 368, 844
\bibitem[Yi et al.(2014)]{2014ApJ...795L..29Y} Yi, W.-M., Wang, F., Wu, X.-B., et al.\ 2014, \apjl, 795, L29
\bibitem[Yuan et al.(2015)]{2015arXiv150607735Y} Yuan, W., Zhang, C., Feng, H., et al.\ 2015, arXiv:1506.07735
\bibitem[Zeng et al.(2013)]{2013MNRAS.431..997Z} Zeng, H., Yan, D., \& Zhang, L.\ 2013, \mnras, 431, 997
\end{thebibliography}
\end{document}